\documentclass[a4paper]{elsarticle} 
\pdfoutput=1 

\journal{Operations Research Letters}

\usepackage{amsmath,amssymb,hyperref,setspace,graphicx,soul} 

\usepackage{natbib}
\usepackage[left=1in,right=1in,top=1.4in,bottom=1.4in]{geometry}
\hypersetup{
	pdfauthor={Jaehyuk Choi},
	colorlinks=true,
	linkcolor=red,
	citecolor=blue,
	urlcolor=blue,
	bookmarksnumbered=true,
	pdfstartview=
}

\date{January 13, 2025}

\newcommand{\vov}{\xi}
\newcommand{\mr}{\kappa}
\newcommand{\vinf}{\theta}
\newcommand{\KL}{Karhunen--Lo\`eve}

\newcommand{\sighat}{\hat{\sigma}}
\newcommand{\Umth}{\bar{U}_{0,T}}
\newcommand{\Vmth}{\bar{V}_{0,T}}
\newcommand{\sigmth}{\bar{\sigma}}
\newcommand{\Usig}{U_{0,T}}
\newcommand{\Vsig}{V_{0,T}}
\newcommand{\even}[1]{\ddot{#1}}
\newcommand{\odd}[1]{\dot{#1}}

\newcommand{\ns}{L}

\newcommand{\qtext}[2][\quad]{#1\text{#2}#1}

\begin{document}
\begin{frontmatter}
\title{Exact simulation scheme for the Ornstein--Uhlenbeck driven stochastic volatility model with the Karhunen--Lo\`eve expansions}

\author[mafn]{Jaehyuk Choi\corref{corrauthor}}
\ead{jc6569@columbia.edu}

\cortext[corrauthor]{Correspondence. \textit{Tel:} +1-212-854-2643, \textit{Address:} Rm 426 Mathematics Hall, 2990 Broadway, New York, NY 10027}

\address[mafn]{Department of Mathematics, Columbia University}

\begin{abstract}
This study proposes a fast exact simulation scheme for the Ornstein--Uhlenbeck driven stochastic volatility model. With the Karhunen--Lo\`eve expansions, the stochastic volatility path (Ornstein--Uhlenbeck process) is expressed as a sine series, and the time integrals of volatility and variance are analytically derived as infinite series of independent normal random variables. The new method is several hundred times faster than the existing method using numerical transform inversion. The simulation variance is further reduced with conditional simulation and the control variate.
\end{abstract}
\end{frontmatter}

\section{Introduction} \noindent
The Ornstein--Uhlenbeck (OU) process has been widely used in quantitative finance as it balances the random walk and mean reversion. In particular, the Ornstein--Uhlenbeck driven stochastic volatility (OUSV) model~\citep{scott1987option,stein1991sv,schobel1999stochastic} is a popular stochastic volatility model for pricing options that exhibits volatility smile. While European options can be priced with the inverse Fourier transform~\citep{schobel1999stochastic}, Monte--Carlo (MC) simulation methods are required for pricing path-dependent derivatives. As such, efficient simulation has been a recent topic of research. Notably, \citet{li2019exact_ou} proposed a novel exact simulation scheme for the OUSV model, contributing to the growing literature on the exact or almost-exact simulations of stochastic volatility models~\citep{broadie2006mcheston,baldeaux2012exact,kang2017exact,cai2017sabr,choi2019nsvh,cui2021efficient,choi2024heston}. The scheme of \citep{li2019exact_ou} made it possible to simulate the asset price after time step of arbitrary size without time discretization. However, the numerical Fourier transform inversion that the scheme relies on is still a computational burden, making it difficult to be widely adopted in practice. According to \citep[Figures~2 and 3]{li2019exact_ou}, their exact scheme starts to outperform the Euler discretization scheme only when the simulation time goes beyond several minutes.

This study proposes a faster exact simulation method for the OUSV model using the \KL{} (KL) expansions of the OU bridge process~\citep{daniluk2016approx}. With the KL expansions, the stochastic volatility path can be represented as an infinite sine series where the coefficients are independent normal random variables. Therefore, it is possible to analytically obtain the time integrals of volatility and variance, which are essential components for the exact simulation. Since it avoids the costly inverse Fourier transform, the new method is several hundred times faster than \citep{li2019exact_ou}. 

The contribution of this study to the OUSV model simulation is compared to those of \citep{glasserman2011gamma,choi2024heston} to the \citet{heston1993closed} model simulation. By expressing the integrated variance under the Heston model with the infinite series of gamma random variables, \citep{glasserman2011gamma} circumvented the computationally expensive transform inversion in the original exact simulation scheme of \citep{broadie2006mcheston}. The gamma expansion algorithm has been further simplified by \citep{choi2024heston}.

The remainder of this paper is organized as follows. In Section~\ref{s:model}, we introduce the OUSV model and its analytical properties. In Section~\ref{s:kl}, we express the volatility path with the KL expansions. In Section~\ref{s:new} and \ref{s:num}, we present the new simulation scheme and its numerical tests, respectively. Finally, we conclude this paper in Section~\ref{s:conc}.

\section{The OUSV model and preliminaries} \label{s:model} \noindent
The stochastic differential equations for the price $S_t$ and the stochastic volatility $\sigma_t$ under the OUSV model are given by
\begin{gather}
	\frac{d S_t}{S_t} = r\, d t + \sigma_t \left( \rho\, d Z_t + \sqrt{1-\rho^2}\, d W_t \right),\\ 
	d \sigma_t = \mr(\vinf - \sigma_t) d t + \vov\, d Z_t, \label{eq:sde_vol}
\end{gather}
where $W_t$ and $Z_t$ are two independent standard Brownian motions, $\rho$ is the correlation between the price and volatility, $\mr$ is the mean reversion speed, $\vov$ is the volatility of volatility, $\vinf$ is the long-term equilibrium volatility, and $r$ is the risk-free rate.

\subsection{Integrated volatility and variance} \noindent
Let $\Usig$ and $\Vsig$ be the time averages of $\sigma_t$ and $\sigma_t^2$, respectively, between $t=0$ and $T$:
\begin{equation} \label{eq:UVsigma}
\Usig = \frac1T \int_0^T \sigma_t \,d t \qtext{and} \Vsig = \frac1T\int_0^T \sigma_t^2 \,d t.
\end{equation}
They are critical state variables for the OUSV model simulation because the terminal price $S_T$, conditional on the triplet ($\sigma_T$, $\Usig$, $\Vsig$), follows a log-normal distribution:
\begin{equation}\label{eq:logreturn}
\log(S_T / S_0) \sim N(\mu_{0,T},\Sigma_{0,T}^2),
\end{equation}
where the mean and variance are given by
\begin{equation*}
	\mu_{0,T} = rT + \frac{\rho}{2\vov} \left[\left(-\vov^2 - 2\mr\vinf \Usig + \left( 2\mr - \vov/\rho\right) \Vsig\right)T + (\sigma_T^2 - \sigma_0^2)\right] \qtext{and} \Sigma_{0,T}^2 = (1-\rho^2)T\Vsig.
\end{equation*}
See \citep[Proposition~1]{li2019exact_ou} for the derivation.
Equivalently, the asset price $S_T$ can be expressed and simulated by a geometric Brownian motion:
\begin{equation}
	\label{eq:S_T}
	S_T \sim F_T \exp\left(\Sigma_{0,T} Z - \frac12 \Sigma_{0,T}^2\right)
\end{equation} 
where $Z$ is a standard normal variable
and $F_T$ is the forward price conditional on ($\sigma_T$, $\Usig$, $\Vsig$):
\begin{equation} \label{eq:condfwd}
\begin{aligned} 
	F_T &= E\{S_T\,|\,\sigma_T, \Usig, \Vsig\} = S_0 \exp\left(\mu_{0,T} + \frac12 \Sigma_{0,T}^2\right)\\
	&= S_0 \exp \left(rT+\frac{\rho}{2\vov} \left[\left(-\vov^2 - 2\mr\vinf \Usig + \left( 2\mr - \rho\vov\right) \Vsig\right)T + (\sigma_T^2 - \sigma_0^2)\right]\right).
\end{aligned}
\end{equation}
Therefore, sampling $S_T$ is reduced to sampling $\sigma_T$, $\Usig$ and $\Vsig$, and the challenge in the OUSV model simulation lies in sampling the triplet. In \citep{li2019exact_ou}, $\sigma_T$ and $\Usig$ are sampled simultaneously as they follow a bivariate normal distribution. Then, $\Vsig$ is drawn from the cumulative distribution function (CDF) obtained from the inverse Fourier transform of $\Vsig$ for given $\sigma_T$ and $\Usig$. Numerical inversion requires a heavy computation. Alternatively, we will adopt the KL expansions to sample $\Usig$, and $\Vsig$ simultaneously for given  $\sigma_T$ (see Section~\ref{s:kl}).

\subsection{Auxiliary processes of the stochastic volatility} \noindent
We define three auxiliary processes of $\sigma_t$ for simulation. First, let us define $\sigmth_t$ and its time averages by removing from $\sigma_t$ the long-term volatility $\vinf$:  
 \begin{equation} \label{eq:sde_x}
	\sigmth_t = \sigma_t - \vinf, \quad
	\Umth = \frac1T \int_0^T \sigmth_t \,d t \qtext{and} \Vmth = \frac1T\int_0^T \sigmth_t^2 \,d t.
\end{equation}
Consequently, the process $\sigmth_t$ satisfies a simpler form of the OU process than Eq.~\eqref{eq:sde_vol}:
$$ 	d \sigmth_t = -\mr \sigmth_t\,d t + \vov \,d Z_t \quad (\sigmth_0 = \sigma_0 - \vinf).
$$
Its solution is well-known as~\citep[Ch.~9]{steele2001stochastic}
$$ \sigmth_T = \sigmth_0 e^{-\mr T} + \vov e^{-\mr T} \int_0^T e^{\mr t} \,d Z_t.
$$
The introduction of ($\sigmth_T$, $\Umth$, $\Vmth$) will simplify algebra. The original triplets ($\sigma_T$, $\Usig$, $\Vsig$) and the new triplet ($\sigmth_T$, $\Umth$, $\Vmth$) are interchangeable by
\begin{equation} \label{eq:conv}
\sigma_T = \vinf + \sigmth_T, \quad
\Usig = \vinf + \Umth, \qtext{and}
\Vsig = \vinf^2 + 2\vinf \Umth + \Vmth.
\end{equation}

Second, we define the centered OU process $\sighat_T$ by removing from $\sigmth_T$ its mean $E(\sigmth_T) = \sigmth_0 e^{-\mr T}$,
$$ \sighat_T = \sigmth_T - \sigmth_0 e^{-\mr T} = \vov e^{-\mr T} \int_0^T e^{\mr t} \,d Z_t \quad (\sighat_0=0).
$$
The centered process $\sighat_T$ is a Gaussian process with zero mean and the covariance given by
$$ \text{Cov}(\sighat_t, \sighat_T) = \frac{\vov^2}{2\mr}\left(e^{-\mr(T-t)} - e^{-\mr(T+t)}\right) = \frac{\vov^2}{\mr} e^{-\mr T} \sinh(\mr t) \qtext{for} 0\le t\le T.
$$
The terminal value $\sighat_T$ can be sampled by
\begin{equation} \label{eq:sigma_T}
	\sighat_T \sim \vov\sqrt{\frac{1-e^{-2\mr T}}{2\mr}}\, Z_0 = \vov \sqrt{T \phi(2\mr T)}\, Z_0 \qtext{for a standard normal $Z_0$,} 
\end{equation}
where $\phi(x)$ is introduced for concise notations:
$$ \phi(x) = \frac{1-e^{-x}}{x}\quad(\phi(0)=1).$$
The terminal value $\sighat_T$ will be interchangeably used with $\sigma_T$ or $\sigmth_T$ via $\sigma_T - \vinf = \sigmth_T = \sighat_T + \sigmth_0 e^{-\mr T}$.

Last, we construct the OU bridge process $B_t$ of $\sighat_t$ ($0\le t \le T$) given the terminal value $\sighat_T$:
$$ B_t = \sighat_t - \frac{\text{Cov}(\sighat_t, \sighat_T)}{\text{Cov}(\sighat_T, \sighat_T)} \sighat_T
= \sighat_t - \frac{\sinh(\mr t)}{\sinh(\mr T)}\sighat_T \quad (B_0=B_T=0),
$$
whose covariance follows as
$$ \text{Cov}(B_s, B_t) = \text{Cov}(\sighat_s, \sighat_t) - \frac{\text{Cov}(\sighat_s, \sighat_T)\,\text{Cov}(\sighat_t, \sighat_T)}{\text{Cov}(\sighat_T, \sighat_T)}.
$$

\section{The KL expansions and its time integrals of the OU process}\noindent
\label{s:kl}
We will express the path of $\sigma_t$ (i.e., OU process) with infinite sine series with the KL expansions. \citet[Theorem 2.3]{daniluk2016approx} has shown that the OU bridge process $B_t$ admits the following KL expansions:
\begin{equation} \label{eq:kl}
B_t = \vov \sum_{n=1}^{\infty} a_n\sqrt{T}\, \sin\left(\frac{n\pi t}{T}\right) Z_n \qtext{for} 0\le t\le T \qtext{and}
a_n = \sqrt\frac{2}{(\mr T)^2 + (n \pi)^2},
\end{equation}
where $Z_n$ are independent standard normal variables. The KL expansions are understood as the principal component analysis (PCA) of $B_t$ in an $\infty$-dimensional space; 
$\sin(n\pi t/T)$ and $a_n\sqrt{T}$ are the eigenfunctions and eigenvalues, respectively, of the covariance function, $\text{Cov}(B_s, B_t)$.

\begin{figure}[ht!]
	\caption{Two sample volatility (OU process) paths generated by KL expansions in Eq.~\eqref{eq:sigma} with $N=2$, 8, 16, and 64 sine terms. The parameters used are 
		$\sigma_0=\vinf = \vov = 0.2$, and $\mr=1$.\label{fig:volpath}}
	\begin{center}
		\includegraphics[width=0.6\textwidth]{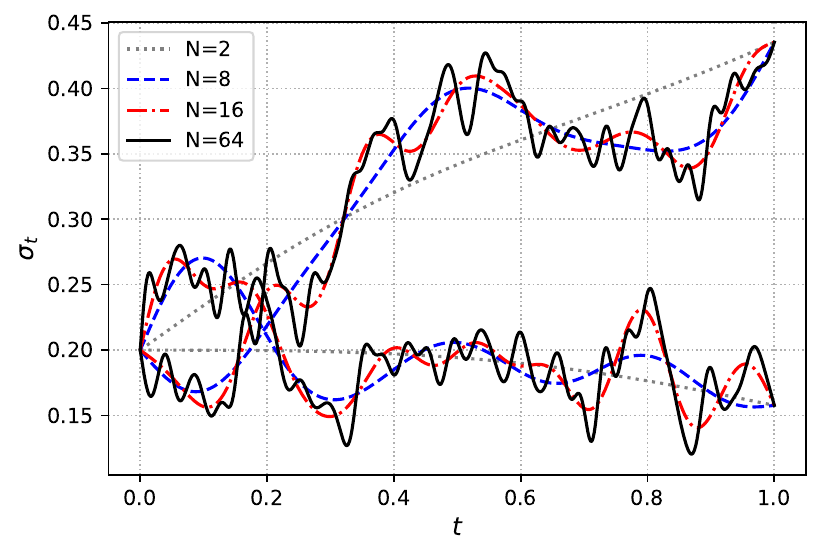}\hspace{2em}
	\end{center}
\end{figure}

Notice that, if $\mr=0$ and $\vov=1$, the centered process $\sighat_t$ (and $\sigmth_t$) is reduced to the standard Brownian motion $Z_t$, and its bridge $B_t$ to the standard Brownian bridge. Accordingly, Eq.~\eqref{eq:kl} is reduced to the well-known KL expansions for the Brownian bridge~\citep{loeve1978probability}: 
$$ B_t = \sum_{n=1}^{\infty} \frac{\sqrt{2T}}{n \pi}\, \sin\left(\frac{n\pi t}{T}\right) Z_n \qtext{for} 0\le t\le T.
$$
Using the KL expansion of $B_t$, the path of $\sigmth_t$, conditional on the terminal value $\sighat_T$, is represented as follows:
\begin{equation} \label{eq:sigma}
\sigmth_t = \sigmth_0 e^{-\mr t} + \sighat_T \frac{\sinh(\mr t)}{\sinh(\mr T)} + 
\vov\sqrt{T}\, \sum_{n=1}^{\infty} a_n \sin\left(\frac{n\pi t}{T}\right) Z_n.
\end{equation}
Figure~\ref{fig:volpath} displays two example paths of $\sigma_t$ represented via Eq.~\eqref{eq:sigma} with the terminal value $\sigma_T$ fixed. As the number of sine terms increases, the paths look closer to those of the OU process.

Since the path of $\sigmth_t$ is decomposed into elementary functions of time, it is now possible to derive $\Umth$ and $\Vmth$ in Eq.~\eqref{eq:sde_x} analytically as infinite series of normal random variables:
\begin{equation} \label{eq:UV}
	\begin{aligned} 
	\Umth =& \underbrace{\left[\sigmth_0 + \frac{\sighat_T}{1+e^{-\mr T}}\right] \phi(\mr T)}_{:=E(\Umth\,|\,\sighat_T)}
	+ 2\vov \sqrt{T} \sum_{\substack{n=1\\n:\text{ odd}}}^\infty \frac{a_n}{n\pi} Z_n, \\
	\Vmth =& \underbrace{\sigmth_0^2\, \phi(2\mr T) + \sighat_T^2\frac{\sinh(2\mr T)-2\mr T}{4\mr T \sinh^2(\mr T)} + \frac{\vov^2}{2\mr}\left[\coth(\mr T) - \frac{1}{\mr T} \right] + \sigmth_0 \sighat_T
	\frac{e^{-\mr T}}{\mr T}\left[\frac{1}{\phi(2\mr T)} - 1\right]}_{:=E\left(\Vmth\,|\,\sighat_T\right)}\\
&+ \vov\sqrt{T} \left[\sigmth_0 \sum_{n=1}^{\infty}n\pi\,a_n^3 Z_n + \sigmth_T \sum_{n=1}^{\infty}(-1)^{n-1}n\pi\,a_n^3 Z_n\right] + \frac{\vov^2 T}{2}\sum_{n=1}^{\infty} a_n^2 (Z_n^2-1).
	\end{aligned}
\end{equation}
See \ref{apdx:trigo} for the detailed derivation, where the trigonometric and hyperbolic integrals are used.

In Eq.~\eqref{eq:UV}, we have easily identified the conditional means, $E(\Umth\,|\,\sighat_T)$ and $E(\Vmth\,|\,\sighat_T)$, because the random variables (i.e., $Z_n$ and $Z_n^2-1$) have zero means.
This is an alternative derivation of \citep[Eq.~(7)]{li2019exact_ou}.
From the conditional means, the unconditional means can also be obtained:
\begin{align*}
E\left(\Usig\right) &= \vinf + (\sigma_0 - \vinf) \phi(\mr T) \\  
E\left(\Vsig\right) &= \vinf^2 + \frac{\vov^2}{2\mr} + 2\vinf (\sigma_0 - \vinf) \phi(\mr T) + 
\left[(\sigma_0 - \vinf)^2 - \frac{\vov^2}{2\mr}\right] \phi(2\mr T) 
\end{align*}
Here, $E\left(\Vsig\right)$ is the fair strike of the continuous variance swap, consistent with \citep[Eq.~(24)]{schobel1999stochastic} and \citep[Eq.~(17)]{bernard2014prices}. Therefore, we have provided an alternative derivation for the fair strike.

\section{New simulation schemes based on the KL expansion} \label{s:new} \noindent
Based on the series expansion of $\Umth$ and $\Vmth$ in Eq.~\eqref{eq:UV}, we propose a new exact simulation scheme. 
In the numerical implementation, we must take only finitely many terms. However, we must not simply ignore the truncated terms as the contribution of them is more important as we take less terms. Therefore, we will sample random variables from the approximated distributions of them to compensate the truncated terms.

Let an even integer $\ns$ be the number of terms to take from the infinite series. The truncated terms in $\Umth$ and $\Vmth$ beyond the first $L$ terms are denoted by
$$ G_\ns = \sum_{\substack{n=\ns+1\\n:\text{ odd}}}^\infty \frac{a_n}{n\pi} Z_n, \quad
P_\ns = \sum_{\substack{n=\ns+1\\n:\text{ odd}}}^{\infty}n\pi\,a_n^3 Z_n, \quad Q_\ns = \sum_{\substack{n=\ns+1\\n:\text{ even}}}^{\infty}n\pi\,a_n^3 Z_n, \qtext{and}
R_\ns = \sum_{n=\ns+1}^{\infty} a_n^2 (Z_n^2-1).
$$
Let us also define several infinite series that are useful in analyzing the above terms:
\begin{equation} \label{eq:bcdfg}
	c_\ns = \sum_{n=\ns+1}^{\infty}\!\! a_n^4, \quad
	f_\ns = \sum_{n=\ns+1}^{\infty} \frac{a_n^2}{(n\pi)^2}, \qtext{and}
	g_\ns = \sum_{n=\ns+1}^{\infty}\!\! (n\pi)^2 a_n^6. 
\end{equation}
The partial sums over the odd and even indices are denoted by single dot and double dots, respectively. For example, 
\begin{equation} \label{eq:oddeven}
\odd{c}_\ns = \sum_{\substack{n=\ns+1\\n:\text{ odd}}}^{\infty}\!\! a_n^4 \qtext{and}
\even{c}_\ns = \sum_{\substack{n=\ns+1\\n:\text{ even}}}^{\infty}\!\! a_n^4.
\end{equation}
The infinite series in Eqs.~\eqref{eq:bcdfg} and \eqref{eq:oddeven} admit analytic expressions (see \ref{apdx:series}) and we will take advantage of them.

We first note that $G_\ns$, $P_\ns$, and $Q_\ns$ follow a multivariate normal distribution with zero means and a covariance matrix:
\begin{equation} \label{eq:cov}
\text{Cov}(G_\ns, P_\ns, Q_\ns) = \begin{pmatrix}
	\odd{f}_\ns & \odd{c}_\ns & 0 \\ \odd{c}_\ns & \odd{g}_\ns & 0 \\ 0 & 0 & \even{g}_\ns
\end{pmatrix}.
\end{equation}
Therefore, we can exactly sample $G_\ns$, $P_\ns$, and $Q_\ns$ by:
\begin{equation}\label{eq:GPQ}
\begin{gathered}
	G_\ns = \sqrt{\odd{f}_\ns}\, \left(\sqrt{1-\rho_\ns^2}\, W_1 + \rho_\ns\, W_2\right) \qtext{with}
	\rho_\ns = \odd{c}_\ns \big/ \sqrt{\odd{f}_\ns\, \odd{g}_\ns}\;,\\
	P_\ns = \sqrt{\odd{g}_\ns}\, W_2 \qtext{and} 
	Q_\ns = \sqrt{\even{g}_\ns}\, W_3, \\
\end{gathered}
\end{equation}
where $W_1$, $W_2$, and $W_3$ are standard normal random variables independent with each other and from $Z_{n\ge 0}$.

Next, we analyze $R_\ns$. The exact distribution of $R_\ns$ is unknown, but the mean and variance of $R_\ns$ can be obtained as
$$ E(R_\ns) = 0 \qtext{and} \text{Var}(R_\ns) = 2 c_\ns.
$$
We cannot say that $R_\ns$ is independent of $G_\ns$, $P_\ns$, and $Q_\ns$ because they are implicitly related via $Z_{n\ge L+1}$. Nevertheless, $R_\ns$ has zero correlation with $G_\ns$, $P_\ns$, and $Q_\ns$ because $E(Z_i(Z_j^2-1))=0$ for all $i$ and $j \ge \ns+1$. Therefore, it is reasonable to approximate $R_\ns$ as
\begin{equation}\label{eq:R}
	R_\ns \approx \sqrt{c_\ns}\,(W_4^2 - 1),
\end{equation}
so that the mean and variance are matched. Here, $W_4$ is a normal random variable independent of $W_1$, $W_2$, and $W_3$.

Finally, the procedure for simulating $\sigmth_T$, $\Umth$, and $\Vmth$ is summarized as
\begin{equation}
\begin{aligned}
	\sigmth_T =& \sigmth_0 e^{-\mr T} + \sighat_T \qtext{where} \sighat_T = \vov\sqrt{T\phi(2\mr T)}\,Z_0,\\
	\Umth =& E(\Umth\,|\,\sighat_T)
	+ 2\vov \sqrt{T} \left(\sum_{\substack{n=1\\n:\text{ odd}}}^\ns \frac{a_n}{n\pi} Z_n + G_\ns \right), \\
	\Vmth \approx& E(\Vmth|\sighat_T) + \vov \sqrt{T} \left[\sigmth_0 \left(\sum_{n=1}^\ns n\pi\,a_n^3 Z_n + P_\ns + Q_\ns\right)\right. \\
	&\qquad \left. +\sigmth_T \left(\sum_{n=1}^\ns(-1)^{n-1}n\pi\,a_n^3 Z_n + P_\ns - Q_\ns\right)\right]
	+ \frac{\vov^2 T}{2}\left(\sum_{n=1}^\ns a_n^2 (Z_n^2-1)+ R_\ns\right),
\end{aligned}
\end{equation}
where $E(\Umth\,|\,\sighat_T)$ and $E(\Vmth|\sighat_T)$ are given in Eq.~\eqref{eq:UV} and $G_\ns$, $P_\ns$, $Q_\ns$, and $R_\ns$ are sampled according to Eqs.~\eqref{eq:GPQ}--\eqref{eq:R}. Once ($\sigmth_T$, $\Umth$, $\Vmth$) is obtained, they can be converted to ($\sigma_T$, $\Usig$, $\Vsig$) by Eq.~\eqref{eq:conv}. Then, $S_T$ can be sampled according to Eqs.~\eqref{eq:S_T}--\eqref{eq:condfwd}.

Several comments on the new algorithm are in order. First, we expect a significant speed gain in our algorithm as it is expressed in terms of elementary functions (e.g., $\sinh$) and standard normal variables only. 
Second, our algorithm contains an approximation in sampling $\Vmth$ only because of $R_\ns$; sampling $\sighat_T$ and $\Umth$ are exact. Third, the number of terms $\ns$ is the only parameter to choose in our algorithm. Whereas in \citep{li2019exact_ou}, there are several parameters to be determined to control the error of the numerical Laplace inversion algorithm.



%

%

\section{Numerical results} \label{s:num} \noindent
We numerically test the performance of the new exact simulation scheme by pricing European options, as in \citep{li2019exact_ou}. Since European options can be priced efficiently by the Fourier inversion~\citep{schobel1999stochastic}, they are not the main target of the MC simulation in practice. Nevertheless, European option serves as a good test case to check the accuracy of the simulation schemes. The availability of a highly accurate benchmark price is an added benefit. 
In the test, we use two additional techniques to minimize the MC variance: conditional MC simulation and martingale-preserving control variate. The reduced variance helps to measure the bias of the simulation scheme more accurately.

While \citep{li2019exact_ou} have priced options by averaging the payoff from the simulated $S_T$ values, we instead price options by averaging the Black--Scholes prices from Eq.~\eqref{eq:S_T} over the simulated values of $(\sigma_T, \Usig, \Vsig)$.
\begin{equation} \label{eq:call}
	\hat{C}_\text{OUSV} = e^{-rT} E_\text{MC}\left\{\,
	C_\text{BS}\!\left(K,\, F_T,\sigma_\text{BS}\right)\,\right\} \qtext{with} \sigma_\text{BS} = \Sigma_{0,T}/\sqrt{T}
\end{equation}
where $E_\text{MC}$ is the MC average and $C_\text{BS}(K, F, \sigma_\text{BS})$ is the undiscounted Black--Scholes price of the call option with strike price $K$, forward price $F$, and volatility $\sigma_\text{BS}$. This method, often called \textit{conditional simulation}~\citep{willard1997condmc}, significantly reduces the MC variance because it uses the theoretical expectation over the asset price $S_T$. The conditional simulation is frequently adopted in the simulation studies~\citep{broadie2006mcheston,cai2017sabr,choi2024heston}. For example, \citep{cai2017sabr} have reported a 99\% reduction in variance under the stochastic alpha-beta-rho (SABR) model.

The accuracy of option price can be further improved with the martingale-preserving control variate on $F_T$. The theoretical mean of the conditional forward price $F_T$ in Eq.~\eqref{eq:condfwd} over $(\sigma_T, \Usig, \Vsig)$ should be the unconditional forward price, $e^{rT}S_0$. However, this is not exactly observed in simulation:
\begin{equation} \label{eq:S0}
	\hat{S}_0 = e^{-rT}\,E_\text{MC}\left\{F_T\right\} \quad (\neq S_0).
\end{equation}
Therefore, we correct $F_T$ by
\begin{equation} \label{eq:fwdcv}
	F_T^\text{cv} = \mu F_T \qtext{for} \mu = S_0\,e^{rT}/E_\text{MC}\left\{F_T\right\},
\end{equation}
and we use $F_T^\text{cv}$ instead of $F_T$ in the Black--Scholes price, Eq.~\eqref{eq:call}.

\begin{table}[!ht]
\caption{The simulation result for $T=1$ and other parameters in Eq.~\eqref{eq:param}. ``Spot Price'' and ``Option Price'' evaluate Eqs.~\eqref{eq:S0} and \eqref{eq:call}, respectively. ``Option Price with CV’’ uses the control variate in Eq.~\eqref{eq:fwdcv}. The true option price is 13.21492. \label{tab:1}}
\begin{center} \small
	\begin{tabular}{|c|r|rr|rr|rr|c|} \hline
		& $n_\text{path}\;\;$ & \multicolumn{2}{c|}{Spot Price} & \multicolumn{2}{c|}{Option Price} & \multicolumn{3}{c|}{Option Price with CV} \\ \cline{3-9}
		$\ns$ & (number & Bias & RMSE & Bias & RMSE & Bias & RMSE & CPU Time \\
		& of paths) & ($\times 10^{-4}$) & ($\times 10^{-2}$) & ($\times 10^{-4}$) & ($\times 10^{-2}$) & ($\times 10^{-4}$) & ($\times 10^{-2}$) & (Seconds) \\ \hline\hline
 & 10,000 & 0.3 & 1.74 & -0.4 & 4.04 & -0.7 & 3.48 & 0.006 \\
2 & 40,000 & 0.3 & 0.87 & -0.4 & 2.33 & -0.6 & 1.74 & 0.024 \\
& 160,000 & 0.3 & 0.44 & -0.4 & 1.17 & -0.6 & 0.88 & 0.096 \\ \hline
& 10,000 & -0.2 & 1.73 & -0.9 & 4.03 & -0.7 & 3.47 & 0.008 \\
4 & 40,000 & -0.2 & 0.87 & -0.9 & 2.33 & -0.7 & 1.74 & 0.028 \\
& 160,000 & -0.2 & 0.44 & -0.9 & 1.17 & -0.7 & 0.87 & 0.109 \\ \hline
& 10,000 & -0.2 & 1.74 & -0.4 & 4.04 & -0.3 & 3.48 & 0.008 \\
6 & 40,000 & -0.2 & 0.87 & -0.4 & 2.33 & -0.2 & 1.74 & 0.030 \\
& 160,000 & -0.2 & 0.44 & -0.4 & 1.17 & -0.2 & 0.87 & 0.122 \\ \hline
	\end{tabular}
\end{center}
\end{table}

In our tests, we measure the spot price $\hat{S}_0$ in Eq.~\eqref{eq:S0} and the option price $\hat{C}_\text{OUSV}$ in Eq.~\eqref{eq:call} with or without the control variate in Eq.~\eqref{eq:fwdcv} for various KL expansion terms ($\ns=2, \cdots, 8$) and MC paths ($n_\text{path}=1,\,4$, and $16 \; \times \; 10^{4}$). We first generate a pool of $256\times 10^{7}$ antithetic samples of $(\sigma_T, \Umth, \Vmth)$ and group them into $n_\text{path} = 1,\,4$, and $16 \; \times \; 10^{4}$ paths, thereby creating $n_\text{set} = 256\times 10^{7} / n_\text{path}$ ($=256, 64$, and $16 \;\times\; 10^{3}$) MC sets. We price $\hat{S}_0$ and $\hat{C}_\text{OUSV}$ within each MC set and report the bias and root mean square error (RMSE) from the $n_\text{set}$ results. We also measure the average CPU time for running one MC set.

For a direct comparison, we use the same parameter set in \citep[Table 2]{li2019exact_ou}:
\begin{equation} \label{eq:param}
	S_0 = K = 100,\; \sigma_0 = \vinf = 0.2,\; \mr = 4,\; \vov=0.1,\; \rho = -0.7, \text{ and } r=0.09531.
\end{equation}
We compute the exact option values for this parameter set with the Fourier inversion, and the prices closely match those reported in \citep{li2019exact_ou}.
\begin{table}[!ht]
	\caption{The simulation result for $T=5$ and other parameters in Eq.~\eqref{eq:param}. ``Spot Price'' and ``Option Price'' evaluate Eqs.~\eqref{eq:S0} and \eqref{eq:call}, respectively. The true option price is 40.79769. ``Option Price with CV'' uses the control variate in Eq.~\eqref{eq:fwdcv}. \label{tab:2}}
	\begin{center} \small
	\begin{tabular}{|c|r|rr|rr|rr|c|} \hline
		& $n_\text{path}\;\;$ & \multicolumn{2}{c|}{Spot Price} & \multicolumn{2}{c|}{Option Price} & \multicolumn{3}{c|}{Option Price with CV} \\ \cline{3-9}
		$\ns$ & (number & Bias & RMSE & Bias & RMSE & Bias & RMSE & CPU Time \\
		& of paths) & ($\times 10^{-4}$) & ($\times 10^{-2}$) & ($\times 10^{-4}$) & ($\times 10^{-2}$) & ($\times 10^{-4}$) & ($\times 10^{-2}$) & (Seconds) \\ \hline\hline
 & 10,000 & 19.7 & 9.56 & 10.7 & 12.13 & -6.8 & 4.11 & 0.006 \\
4 & 40,000 & 19.7 & 4.78 & 10.7 & 6.35 & -6.9 & 2.06 & 0.026 \\
& 160,000 & 19.7 & 2.41 & 10.7 & 3.19 & -7.0 & 1.04 & 0.103 \\ \hline
& 10,000 & 5.4 & 9.57 & 3.4 & 12.13 & -1.2 & 4.12 & 0.007 \\
6 & 40,000 & 5.4 & 4.79 & 3.4 & 6.37 & -1.4 & 2.06 & 0.028 \\
& 160,000 & 5.4 & 2.39 & 3.4 & 3.18 & -1.4 & 1.03 & 0.110 \\ \hline
& 10,000 & -0.2 & 9.54 & -1.2 & 12.11 & -0.7 & 4.11 & 0.007 \\
8 & 40,000 & -0.2 & 4.76 & -1.2 & 6.33 & -0.9 & 2.05 & 0.028 \\
& 160,000 & -0.2 & 2.37 & -1.2 & 3.15 & -0.9 & 1.02 & 0.111 \\ \hline
	\end{tabular}
\end{center}
\end{table}

Tables~\ref{tab:1}, \ref{tab:2}, and \ref{tab:3} show the results for the maturities, $T=1$, 5 and 10, respectively. As expected, the results become more accurate as $L$ increases. Overall, single-digit $L$ is enough to yield very small bias in the order of $10^{-4}$. The martingale-preserving control variate on $F_T$ further reduces both bias and RMSE simultaneously. With the control variate, the new simulation scheme produces accurate option prices even with $\ns=4$. Most importantly, the new simulation method is several hundred times faster than the CPU time reported in \citep{li2019exact_ou}. This study implemented the simulation in Python on a personal computer running the Windows 10 operating system with an Intel Core i5--6500 (3.2 GHz) CPU and 8 GB RAM whereas \citep{li2019exact_ou} used Intel Core i5--4200U (2.29 GHz) CPU.


\begin{table}[!ht]
\caption{The simulation result for $T=10$ and other parameters in Eq.~\eqref{eq:param}. ``Spot Price'' and ``Option Price'' evaluate Eqs.~\eqref{eq:S0} and \eqref{eq:call}, respectively. The true option price is 62.76312. ``Option Price with CV'' uses the control variate in Eq.~\eqref{eq:fwdcv}.\label{tab:3}}
\begin{center} \small
	\begin{tabular}{|c|r|rr|rr|rr|c|} \hline
		& $n_\text{path}\;\;$ & \multicolumn{2}{c|}{Spot Price} & \multicolumn{2}{c|}{Option Price} & \multicolumn{3}{c|}{Option Price with CV} \\ \cline{3-9}
		$\ns$ & (number & Bias & RMSE & Bias & RMSE & Bias & RMSE & CPU Time \\
		& of paths) & ($\times 10^{-4}$) & ($\times 10^{-2}$) & ($\times 10^{-4}$) & ($\times 10^{-2}$) & ($\times 10^{-4}$) & ($\times 10^{-2}$) & (Seconds) \\ \hline\hline
 & 10,000 & 19.3 & 19.14 & 16.8 & 20.68 & -1.3 & 2.65 & 0.007 \\
6 & 40,000 & 19.3 & 9.59 & 16.8 & 10.56 & -1.6 & 1.33 & 0.028 \\
& 160,000 & 19.3 & 4.78 & 16.8 & 5.27 & -1.7 & 0.66 & 0.111 \\ \hline
& 10,000 & -1.5 & 19.06 & -1.9 & 20.63 & 0.0 & 2.64 & 0.007 \\
8 & 40,000 & -1.5 & 9.50 & -1.9 & 10.47 & -0.3 & 1.32 & 0.029 \\
& 160,000 & -1.5 & 4.73 & -1.9 & 5.20 & -0.4 & 0.65 & 0.117 \\ \hline
& 10,000 & -4.6 & 19.10 & -4.5 & 20.65 & 0.5 & 2.65 & 0.008 \\
10 & 40,000 & -4.6 & 9.57 & -4.5 & 10.54 & 0.2 & 1.33 & 0.033 \\
& 160,000 & -4.6 & 4.82 & -4.5 & 5.31 & 0.1 & 0.67 & 0.132 \\ \hline
	\end{tabular}
\end{center}
\end{table}


\section{Concluding remarks} \label{s:conc}\noindent
This study approximates the OU process (i.e., the volatility process) with the KL expansions composed of a sine series. The KL expansions enable analytical computation of the integrals of volatility and variance, significantly speeding up the exact simulation of the OUSV model. This study is a new addition to the stream of research that applies the KL expansions to quantitative finance areas, such as volatility surface~\citep{cont2002dynamics} and interest rate term-structure models~\citep{daniluk2016approx}. We hope to see more applications of the KL expansion to other stochastic volatility models.

\appendix
\section{Integrals of trigonometric and hyperbolic functions} \label{apdx:trigo} \noindent
This appendix provides more detail on the derivation of $\Umth$ and $\Vmth$ in Eq.~\eqref{eq:UV} from the KL expansion, Eq.~\eqref{eq:sigma}. The expression for $\Umth$ is obtained without difficulty based on 
$$ \int_0^1 \sin(n\pi x) dx = \begin{cases} \quad 0 & \text{for even $n$} \\
	2\,/\,n\pi & \text{for odd $n$}. \end{cases}
$$

The expression of $\Vmth$ is first obtained as
\begin{equation} \label{eq:UV_detail}
\begin{aligned}
	\Vmth =& \sigmth_0^2\, \phi(2\mr T) + \sighat_T^2\frac{\sinh(2\mr T)-2\mr T}{4\mr T \sinh^2(\mr T)} + \frac{\vov^2 T}{2}\sum_{n=1}^{\infty} a_n^2 Z_n^2 + \sigmth_0 \sighat_T
	\frac{e^{-\mr T}}{\mr T}\left(\frac{1}{\phi(2\mr T)} - 1 \right)\\
	&+ \sigmth_0\vov \sqrt{T} \sum_{n=1}^{\infty}\left(1+(-1)^{n-1}e^{-\mr T}\right)n\pi\,a_n^3 Z_n + \sighat_T \vov \sqrt{T} \sum_{n=1}^{\infty}(-1)^{n-1}n\pi\,a_n^3 Z_n.
\end{aligned}
\end{equation}
The six terms above are obtained in the order of
$$ \int_0^T \!dt\, (A+B+C)^2 = \int_0^T \!dt\, (A^2 + B^2 + C^2 + 2AB + 2AC + 2BC),
$$ where $A$, $B$, and $C$ are the three terms of $\sigmth_t$ in Eq.~\eqref{eq:sigma}:
$$ 
A:=\sigmth_0 e^{-\mr t}, \quad B:=\sighat_T \frac{\sinh(\mr t)}{\sinh(\mr T)}, \qtext{and}
C:=\vov\sqrt{T}\, \sum_{n=1}^{\infty} a_n \sin\left(\frac{n\pi t}{T}\right) Z_n.
$$
In the integrals of $C^2$, $AC$, and $BC$, the following integrals of the trigonometric and hyperbolic functions have been used respectively:
\begin{gather*}
	\int_{0}^{1} \sin(m\pi x)\sin(n\pi x)dx = \frac12 \delta_{mn},\\
	\int_0^1 e^{-\lambda x}\sin(n\pi x)dx = \frac{n\pi}{\lambda^2+(n\pi)^2}\left(1+(-1)^{n-1}\,e^{-\lambda}\right),\\
	\int_0^1 \sinh(\lambda x)\sin(n\pi x)dx = \frac{(-1)^{n-1}n\pi}{\lambda^2+(n\pi)^2}\sinh(\lambda).
\end{gather*}
where $m$ and $n$ are integer values, and $\delta_{mn}$ is the Kronecker delta.

Using the analytic expressions of infinite series,
\begin{equation} \label{eq:b0}
	b_0(\lambda) = \sum_{n=1}^\infty \frac{2}{\lambda^2+(n\pi)^2}=\frac{\lambda\coth(\lambda)-1}{\lambda^2} \; \left(\frac13 \qtext{if} \lambda=0\right),\\
\end{equation}
we can re-express the third term of \eqref{eq:UV_detail} with $\lambda=\mr T$:
$$ \sum_{n=1}^{\infty} a_n^2 Z_n^2 =  \sum_{n=1}^{\infty} a_n^2 (Z_n^2 - 1) + \frac{\mr T\coth(\mr T) - 1}{(\mr T)^2}.$$
After simplifying the last two terms with $\sigmth_T = \sigmth_0 e^{-\mr T} + \sighat_T$, we obtain the expression of $\Vmth$ presented in Eq.~\eqref{eq:UV}.

In the derivation of $\Umth$ and $\Vmth$ above, we exchanged the order of the infinite sum and integral. This is possible because the KL theorem~\citep{loeve1978probability} states that, given a path of $B_t$ ($0\le t\le T$), the convergence of the KL expansion in Eq.~\eqref{eq:kl} to the path is uniform in $t$. 

\section{Analytic solutions of infinite series} \label{apdx:series}\noindent
The infinite sums in Eqs.~\eqref{eq:bcdfg} can be evaluated analytically. For convenience, we re-define $a_n$ as a function of $\lambda=\mr T$:
$$ a_n = \sqrt{\frac{2}{\lambda^2 + (n\pi)^2}}.
$$
Therefore, $c_\ns$, $f_\ns$, and $g_\ns$ in Eq.~\eqref{eq:bcdfg} are also understood as functions of $\lambda=\mr T$. In addition to the infinite sums in Eq.~\eqref{eq:bcdfg}, we extend the definition of $b_0(\lambda)$ in Eq.~\eqref{eq:b0} and define $d_\ns(\lambda)$ to be used below:
$$ b_\ns(\lambda) = \sum_{n=\ns+1}^\infty a_n^2 \qtext{and} d_\ns(\lambda) = \sum_{n=\ns+1}^{\infty} a_n^6.
$$
The infinite sums, $c_0(\lambda)$ and $d_0(\lambda)$ (i.e., $\ns=0$), are sequentially derived as the derivatives with respect to $\lambda$:
\begin{align*}
	c_0(\lambda) &= \sum_{n=1}^\infty \frac{4}{(\lambda^2+(n\pi)^2)^2} = -\frac{1}{\lambda}\frac{\partial}{\partial \lambda} b_0(\lambda) = \frac{1}{\lambda^4}\left(\frac{\lambda}{\tanh(\lambda)}+\frac{\lambda^2}{\sinh(\lambda)^2} -2\right),\\
	d_0(\lambda) &= \sum_{n=1}^\infty \frac{8}{(\lambda^2+(n\pi)^2)^3} = -\frac{1}{2\lambda}\frac{\partial}{\partial \lambda} c_0(\lambda)
	= \frac{1}{2\lambda^6} \left( \frac{3\lambda}{\tanh(\lambda)} + \frac{\lambda^2(3 + 2\lambda/\tanh(\lambda))}{\sinh(\lambda)^2} -8 \right).
\end{align*}
Here, we interchanged the differentiation and summation because the convergence of $b_0(\lambda)$ and $c_0(\lambda)$ are uniform in $\lambda\ge 0$.

The infinite sums, $f_0(\lambda)$ and $g_0(\lambda)$, are derived as the differences:
\begin{align*}
	f_0(\lambda) &= \sum_{n=1}^\infty \frac{2}{(n\pi)^2(\lambda^2 + (n\pi)^2)} = \frac{1}{\lambda^2}\sum_{n=1}^\infty \left[\frac{2}{(n\pi)^2} - \frac{2}{\lambda^2+(n\pi)^2}\right]
	= \frac{b_0(0)-b_0(\lambda)}{\lambda^2},\\ 
	g_0(\lambda) &= \sum_{n=1}^\infty \frac{8\,(n\pi)^2}{(\lambda^2+(n\pi)^2)^3} = \sum_{n=1}^\infty \left[\frac{8}{(\lambda^2+(n\pi)^2)^2} - \frac{8\lambda^2}{(\lambda^2+(n\pi)^2)^3}\right]
	= 2c_0(\lambda) - \lambda^2 d_0(\lambda).
\end{align*}

The sums of the even- and odd-indexed terms in $c_0$, $d_0$, and $f_0$ can also be derived.
For example, the even-indexed sum in $c_0$ is given by
$$ \even{c}_0(\lambda) 
= \sum_{\substack{n=1\\n:\text{ even}}}^\infty \frac{4}{(\lambda^2+(n\pi)^2)^2}
= \sum_{n=1}^\infty \frac{4}{(\lambda^2+(2n\pi)^2)^2}
= \frac{1}{2^4} \sum_{n=1}^\infty \frac{4}{((\lambda/2)^2+(n\pi)^2)^2} = \frac{c_0(\lambda/2)}{16}.
$$
The similar trick applies to $f_0(\lambda)$ and $g_0(\lambda)$:
$$ \even{f}_0(\lambda) = \frac{f_0(\lambda/2)}{16} \qtext{and}
\even{g}_0(\lambda) = \frac{g_0(\lambda/2)}{16}.
$$
Then, the sums of the odd-indexed terms follow as
\begin{gather*}
\odd{c}_0(\lambda) = c_0(\lambda) - \even{c}_0(\lambda) = c_0(\lambda) - \frac{c_0(\lambda/2)}{16},\\
\odd{f}_0(\lambda) = f_0(\lambda) - \frac{f_0(\lambda/2)}{16}, \qtext{and}
\odd{g}_0(\lambda) = g_0(\lambda) - \frac{g_0(\lambda/2)}{16}.
\end{gather*}
Finally, the sums for $L>0$ are obtained by subtracting the first $\ns$ terms. For example,
$$ c_\ns(\lambda) = c_0(\lambda) - \sum_{n=1}^\ns a_n^4, \quad
\even{c}_\ns(\lambda) = \even{c}_0(\lambda) - \sum_{\substack{n=1\\n: \text{ even}}}^\ns\!\! a_n^4, \qtext{and}
\odd{c}_\ns(\lambda) = \odd{c}_0(\lambda) - \sum_{\substack{n=1\\n:\text{ odd}}}^\ns\!\! a_n^4.
$$
The same applies to $f_\ns$ and $g_\ns$.

\begin{singlespace}
\bibliography{OUSV-Abbr}
\end{singlespace}
\end{document}